# A Tunable Phonon-Exciton Fano System in Bilayer Graphene

Tsung-Ta Tang*[1], Yuanbo Zhang*[1], Cheol-Hwan Park[1], Baisong Geng[1], Caglar Girit[1], Zhao Hao[2,4], Michael C. Martin[2], Alex Zettl[1,3], Michael F. Crommie[1,3], Steven G. Louie[1,3], Y. Ron Shen[1,3] and Feng Wang[†1,3]

[1] Department of Physics, University of California at Berkeley, Berkeley, CA 94720, USA

[2] Advanced Light Source Division, Lawrence Berkeley National Laboratory, Berkeley, CA 94720, USA

[3] Materials Science Division, Lawrence Berkeley National Laboratory, Berkeley, CA 94720, USA

[4] Earth Sciences Division, Lawrence Berkeley National Laboratory, Berkeley, CA 94720, USA

* These authors contributed equally to this work.

[†] To whom correspondence should be addressed. Email: fengwang76@berkeley.edu

**Interference between different possible paths lies at the heart of quantum physics. Such interference between coupled discrete and continuum states of a system can profoundly change its interaction with light as seen in Fano resonance[1,2]. Here we present a unique many-body Fano system composed of a discrete phonon vibration and continuous electron-hole pair transitions in bilayer graphene[3-7]. Mediated by the electron-phonon interactions, the excited state is described by new quanta of elementary excitations of hybrid phonon-exciton nature. Infrared absorption of the hybrid states exhibit characteristic Fano lineshapes with parameters renormalized by many-body interactions. Remarkably, the Fano resonance in bilayer graphene is continuously tunable through electrical gating. Further control of the phonon-exciton coupling may be achieved with an optical field exploiting the excited state infrared activity. This tunable phonon-exciton system also offers the intriguing possibility of a 'phonon laser' with stimulated phonon amplification generated by population inversion of band-edge electrons.**

Two dimensional (2D) graphene exhibits remarkable electrical[3,4], vibrational[8-11], and optical properties[12-16]. Many of these properties can be modified in graphene through electrical gating, which enables control of the intricate interplay between electrons, phonons, and photons. Such control is exemplified in gated bilayer graphene structures, where an gating electrical field can modify *phonon-light* interactions by inducing infrared (IR) activity in an otherwise IR-inactive phonon vibration (Fig. 1a, b), as well as *electron-light* interactions by generating a tunable electronic bandgap[5,16-21] (Fig. 1c, d). In this way *electron-phonon* interactions are also modified through the electronic structure and phonon dipole changes[9-11]. This tunable coupling between electrons, vibrations, and light offers exciting opportunities in exploring new physical phenomena. One example is the control of quantum interference between vibration- and electron-mediated optical absorption (Fig. 1e), which provides an unusual tunable Fano system involving three different elementary excitations: photon, exciton, and phonon.

A Fano resonance describes the quantum interference of optical absorption through coupled discrete and continuum transitions, where the excited eigenstates are mixtures of the discrete and continuum states[1,2]. It gives rise to a characteristic asymmetric lineshapes in absorption spectra[1]. First introduced to describe atomic photoionization[2], the Fano lineshape has been observed ubiquitously in neutron scattering[22], Raman scattering[23], photoabsorption in quantum wells[24], and electrical transport through nanostructures[25,26]. However, a *tunable* Fano system characterized by interference between *many-body excitations* is difficult to achieve. Here bilayer graphene provides a unique many body system where Fano interference between coupled phonon and electron-hole pair excitations can be realized and controlled through electrical gating. Unlike the relatively simple Fano resonances characterized by single-electron transitions in atoms and quantum dots, the bilayer graphene Fano resonance describes a collective phonon vibration coupled to a continuum of electron-hole excitations through electron-phonon interactions. Such coupling leads to new elementary excitations described by hybrid phonon-exciton states, in which the 'dressed' phonon component has a strongly renormalized center frequency, line width, and infrared activity due to the many-body interactions. This hybrid phonon-exciton excitation and the associated tunable Fano resonance in bilayer graphene can be continuously tuned through electrical gating and are described quantitatively using a tight-binding model. This microscopic understanding of the couplings between electron, phonon, and light in bilayer graphene can lead to novel applications exploiting the tunable bandgap of bilayer graphene.

In our studies, we used dual-gated bilayer graphene samples. Fig. 1f shows an optical microscopy image of a typical device. A pristine A-B stacked graphene bilayer has inversion symmetry, but an electrical field normal to the graphene plane breaks this symmetry and profoundly changes the graphene electronic and vibrational properties. Electronically, bilayer graphene is intrinsically an undoped gapless semiconductor. A normal electrical field, however, induces a non-zero energy gap and charge carrier doping. These two key parameters, electronic bandgap ($\Delta$) and carrier doping concentration (n), can be controlled independently in a dual-gate graphene device. This is achieved by controlling both the top and bottom displacement fields $D_t$ and $D_b$ with the corresponding gates (Fig. 1g) : the discontinuity of the displacement field $\Delta D = D_b - D_t$ determines the carrier doping concentration, while the average displacement field $\bar{D} = (D_b+D_t)/2$ determines the tunable bandgap [16] (see supplementary information). For phonon vibrations, we focus on the symmetric G-mode phonon[6,11] (Fig. 1a). It carries zero dipole moment and is therefore infrared inactive in the pristine symmetric bilayer. A normal electric field, however, breaks the symmetry between the upper and lower layers, resulting in different charges on the carbon atoms at the A site of the bottom layer and the B site of the top layer (Fig. 1b). These atoms now appear like oppositely charged ions in a polar crystal and their vibrations become infrared active.

The field-induced electronic bandgap and phonon infrared activity lead to dramatic changes in graphene bilayer infrared absorption spectra. Fig. 1h displays the absorption spectrum of an ungated bilayer sample, whereas Fig. 1i shows the gate-induced absorption spectrum with carrier doping n ~ 0 and bandgap $\Delta$ ~ 190 meV. While the ungated bilayer absorption spectrum is largely featureless, the spectrum with $\Delta$ ~ 190 meV shows two prominent new features: a broad absorption peak corresponding to electronic bandgap transitions, and a sharp spectral feature at 195 meV coincident with the G-phonon energy. Instead of seeing an absorption peak as might be expected for an IR active phonon, we surprisingly observe a strong absorption dip at the phonon frequency. This 'anti-resonance' signifies that the phonon and electronic excitations are not independent in bilayer graphene. Instead, they form new elementary excitation of hybrid phonon-exciton nature. Such new elementary excitations strongly modify light absorption around the phonon resonance. It can be described as a many-body realization of the Fano interference, where the discrete state is the phonon vibration and the continuum states are the electron-hole pair excitations.

Phenomenologically a Fano resonance absorption lineshape is described by

$$A(E) = A_e \cdot \frac{[q \cdot \gamma + (E - E_\Omega)]^2}{(E - E_\Omega)^2 + \gamma^2}$$ 1, where $A_e$ is the bare electronic state absorption, and $E_\Omega$ and $\gamma$ are the center frequency and width of the phonon resonance, respectively. In the numerator, $\gamma$ and $(E - E_\Omega)$ describe the phonon and electronic weights in the hybrid wavefunction at different energies, and the dimensionless parameter q characterizes the relative dipole strength of the renormalized phonon and electron transitions. Depending on the value of q, the absorption can assume resonance ($|q|\gg 1$, phonon dominates), dispersive ($|q|\sim 1$, comparable phonon and electron contribution), or antiresonance ($|q|\ll 1$, electrons dominate) lineshapes. In bilayer graphene we can achieve these different lineshapes by controlling the electron and phonon transition dipoles with different gate electrical fields.

We show in Fig. 2a the Fano resonance absorption at different bandgap energies (Δ) and zero electron doping (black lines). To focus on the sharp Fano features, we have subtracted the broad electronic absorption background, which was approximated by a third order polynomial fitting of the absorption in the frequency range 150-260 meV excluding the phonon region of 185-215 meV (red dashed line in Fig. 1f). These absorption spectra are well described by the Fano lineshape

$$A(E) - A_e = A_e \cdot \left( \frac{[q \cdot \gamma + (E - E_\Omega)]^2}{(E - E_\Omega)^2 + \gamma^2} - 1 \right),$$ where $A_e$, q, $E_\Omega$, $\gamma$ are the fitting parameters

(fits are shown in Fig. 2a as red traces). In the relatively narrow spectral range around the Fano resonance, $A_e$ has been approximated as a constant. The Fano lineshape is seen to change dramatically when the energy gap Δ is increased, first as a weak absorption dip, then as a dispersive interference, and finally as an absorption peak. The critical parameter describing these lineshapes is q. In these spectra, q is negative and its magnitude increases by an order of magnitude as $\Delta$ increases. This increase of |q| can be qualitatively understood by noting that the phonon gains infrared activity from the field-induced symmetry breaking and its dipole moment increases monotonically with the field strength. At the same time electronic transitions at $E = E_\Omega$ get weaker with increasing field strength once the gap energy is larger than $E_\Omega$.

Instead of varying the electronic bandgap, we can also modify the Fano resonance infrared absorption by changing the carrier concentration. This is achieved experimentally by fixing the bottom gate voltage and sweeping the top gate voltage. In such measurements, the bandgap also changes slightly, but the dominant effects are from carrier doping. Fig. 2b shows Fano absorption spectra (with the broad electronic absorption background subtracted) at different carrier concentrations. We observe that electron doping can also lead to a continuous change from a dip to a peak in the absorption spectrum around the phonon energy. In this case, the increase in |q| is mainly due to a reduction of electronic transition strength from Pauli blocking: with the increase of the electron density, the Fermi energy shifts up and the electronic transitions at $E_\Omega$ become forbidden when the final states for such transitions are already filled.

The detailed dependence of Fano resonance parameters on electrical gating reveals not only the changes of electron and phonon optical dipole strengths contained in $A_e$ and q, but also the renormalization of the 'dressed' phonon vibration frequency $E_\Omega$ and broadening $\gamma$. Displayed in Fig. 3a-3d (symbols) are the experimental data for $A_e$, q, $E_\Omega$, and $\gamma$ at different gap energies with n ~ 0. All of these behaviors of the tunable many-body Fano resonance can be described quantitatively in bilayer graphene by calculating the band structure using atight-binding model and by introducing a hybrid phonon-exciton excitation $h_E^+ = \sin\theta_E \cdot (A \cdot a^+ + \sum_{E' \neq E} B \cdot d_{E'}^+) + \cos\theta_E \cdot d_E^+$. Here $d_E^+$ creates a specific superposition of electron-hole pair states at energy $E$, $a^+$ creates a bare phonon, and A, B and $\theta_E$ are energy dependent parameters (see supplementary information for details) [1]. The first term in the expression describes the renormalized phonon vibration, which is a bare phonon dressed by off-resonance electron-hole pair excitations. It is through this coupling that the phonon gains infrared activity.

We show in Fig. 3 the comparison of experimental values (symbols) and theoretical results (lines). We have used an energy broadening of 40 meV for the electronic transitions to account for the finite excited state lifetime and inhomogeneity in environmental carrier doping from charged impurities. A bare phonon energy of 197.1 meV and width of 0.6 meV in the absence of electron-phonon interactions were used as fitting parameters in Fig. 3c and Fig. 3d, respectively. Our fitting is consistent with tight binding parameters of an intralayer hopping energy of 3 eV, an interlayer hopping energy of 0.4 eV, and vibration induced hopping energy change of 6 eV/Å.[9,27] We note some interesting relations between the Fano parameters. The phonon frequency is red-shifted and q assumes a negative value in this coupled electron-phonon system. These are both due to the fact that the phonon coupling to higher energy continuum states is greater than the coupling to lower energy states. Phonon width ($\gamma$) and electronic absorption (A$_e$) also show a common maximum when the bandgap energy is tuned close to the phonon energy (at 195 meV), both due to a dramatic increase in the electronic density of states. The additional broadening of the phonon width due to electron-phonon coupling reaches 1.4 meV, which is significantly larger than its intrinsic width of 0.6 meV. This corresponds to a maximum phonon decay rate of 4.2 ps$^{-1}$ from electron-hole pair generation and 1.8 ps$^{-1}$ from anharmonic phonon coupling.

In conclusion, gated bilayer graphene exhibits rich Fano resonance behavior. It features coupled resonant light, exciton, and phonon excitations enabled by electric field tuning. In the weak optical excitation limit we observe that the composite exciton-phonon states change infrared light absorption dramatically. If a strong optical excitation is introduced, further control of the hybrid phonon-exciton elementary excitation will be possible. We have also discovered a very strong resonant coupling between the phonon and bandgap electronic transitions in bilayer graphene. It can lead to a 'phonon gain' with a population inversion of electrons. The maximum 'phonon gain' is set by the stimulated phonon emission rate (from fully population inverted electrons), which equals the phonon absorption rate by unexcited electrons and is about 4.2 ps$^{-1}$. In comparison, the 'phonon loss' is determined by the anharmonic phonon decay rate at ~ 1.8 ps$^{-1}$. Therefore a phonon laser with stimulated phonon emission gain overcoming the loss can be achieved in bilayer graphene if sufficient population inversion is created by either optical pumping or electron injection.

**Methods**

The fabrication procedure for creating the dual-gate devices was described in Ref.[16]. Briefly, bilayer graphene was connected to 20nm thick Au source and drain electrodes. The doped Si substrate under a 285 nm thick SiO$_2$ layer served as the bottom gate, and a semi-transparent strip of Pt film on top of an 80 nm thick Al$_2$O$_3$ film formed the top gate.

The Au and Pt electrodes were deposited through stencil masks in vacuum. A cross-sectional view of the bilayer device is sketched in Fig. 1g. Infrared absorption of the Fano resonance was measured using the synchrotron-based IR source from the Lawrence Berkeley National Lab Advanced Light Source and a micro-Fourier transform infrared spectrometer. All experiments were carried out at the room temperature.

**Acknowledgements** This work was supported by University of California at Berkeley and the Office of Basic Energy Sciences, U.S. Department of Energy under contract No. DE-AC03-76SF0098 (Materials Science Division) and contract No. DE-AC02-05CH11231 (Advanced Light Source). Y. Z. and F.W. acknowledge support from a Miller fellowship and a Sloan Fellowship, respectively. T.T.T. is partially supported by National Science Council, Taiwan.

**Author contributions** F.W designed the experiment. T.T.T, Y.Z. B.G. and C.G. fabricated the sample. T.T.T., Y.Z., Z.H., M.C.M., and F.W. performed infrared spectroscopy measurements. C.H.P., S.G.L and F.W. did the calculations. T.T.T., Y.Z., C.H.P, A.Z., M.F.C., S.G.L, Y.R.S. and F.W. wrote the paper together.

The authors declare no competing financial interests.

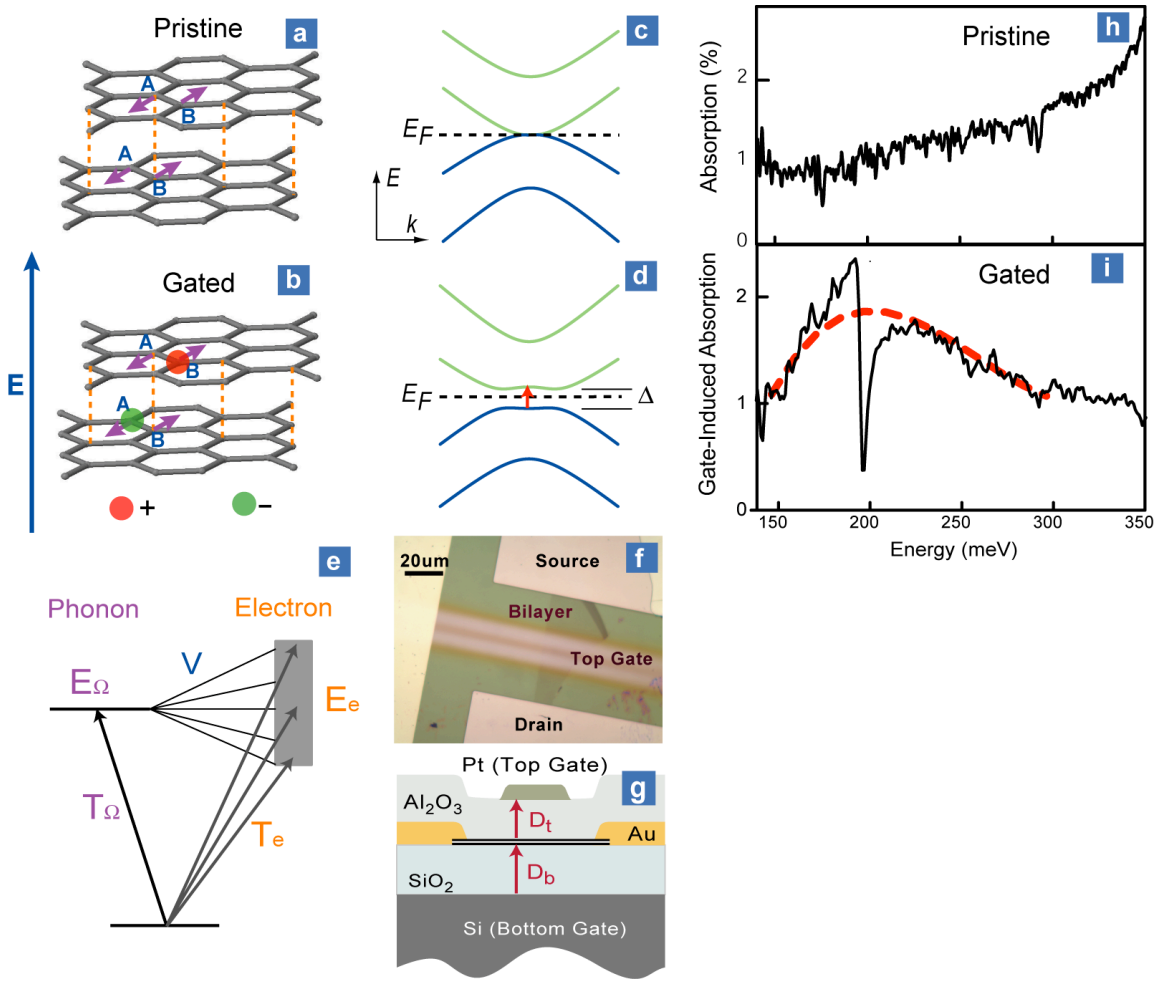

Fig. 1: Dual-gate bilayer graphene as a tunable Fano system. (a) Symmetric optical phonon in a pristine A-B stacking graphene bilayer has zero dipole moment. (b) Field-induced charge redistribution leads to infrared activity in the optical phonon. (c) Pristine bilayer graphene has a zero bandgap. (d) Electrical field generates a finite bandgap in bilayer graphene. (e) A diagram of the Fano resonance describing the interfering infrared transitions to the coupled discrete phonon and continuum electronic states. (f) Optical microscope image of a typical dual-gate graphene device. (g) Illustration of the bilayer device, side view. (h) Infrared absorption from a pristine bilayer. The absorption is relatively flat and no absorption resonance is present in this spectral range. (i) Gate-induced infrared absorption spectrum of the bilayer graphene at zero doping and 190 meV bandgap energy. The broad absorption peak (fitted by a polynomial function in the red dashed curve) arises from bandgap electronic transitions. The sharp spectral feature at 195 meV is due to the optical phonon. The phonon resonance appears as a dip rather than an absorption peak because of Fano interference.

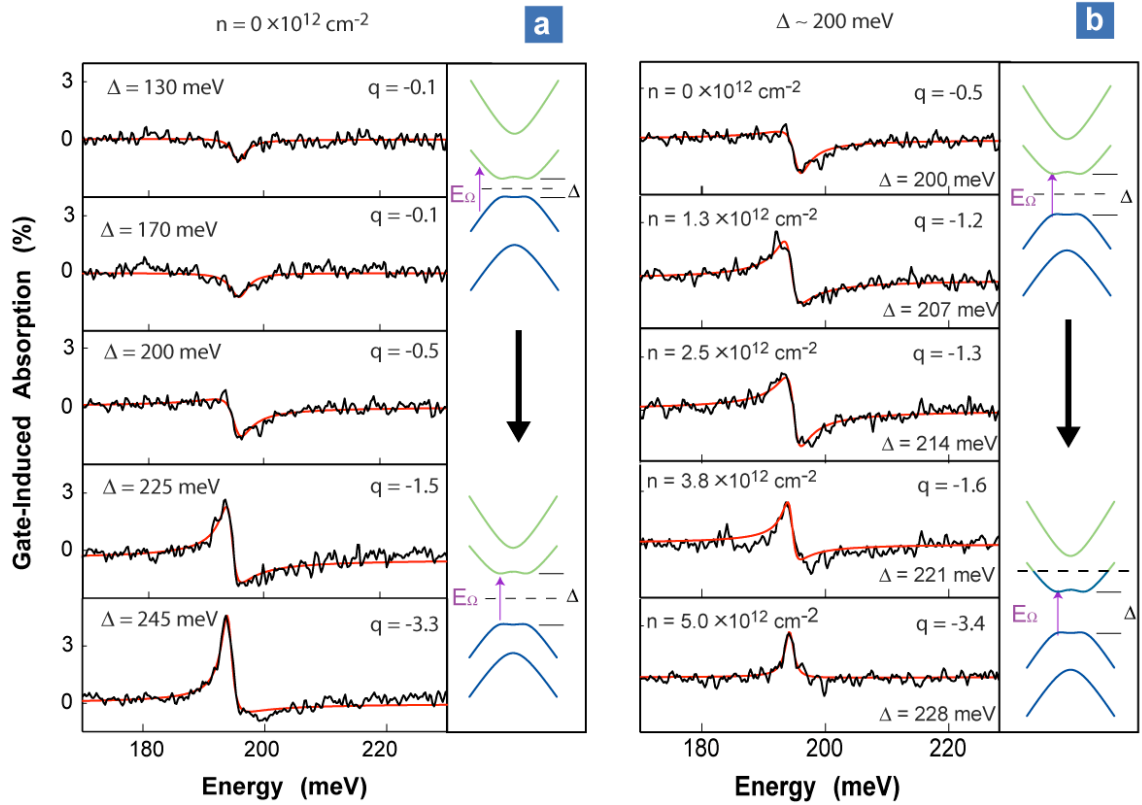

Fig. 2: Infrared absorption spectra of the gate-tunable Fano resonance. (a) Controlling Fano resonance by varying bandgap energy Δ (right panel). From top to bottom Fano resonance absorption spectra show the Fano q-factor increasing with increasing Δ. (Broad electronic absorption backgrounds are subtracted.) Black curves are experimental data and red curves are fits with Fano lineshape. The lineshape changes from an 'anti-resonance' to dispersive-like to a resonance, corresponding to a change of interference parameter |q|<<1 (dominated by electronic absorption with Δ< $E_\Omega$ ) to |q|>>1 (dominated by phonon absorption with Δ > $E_\Omega$). (b) Controlling Fano resonance by increasing electron concentration at Δ ~ 200 meV (right panel). Different Fano lineshapes are observed with increasing electron doping, corresponding to an increase of |q|. This effect is mainly due to a reduction of electronic dipole strength because the final states for

electronic transitions $E_\Omega$ are Pauli blocked by doped electrons, as illustrated in the right panel.

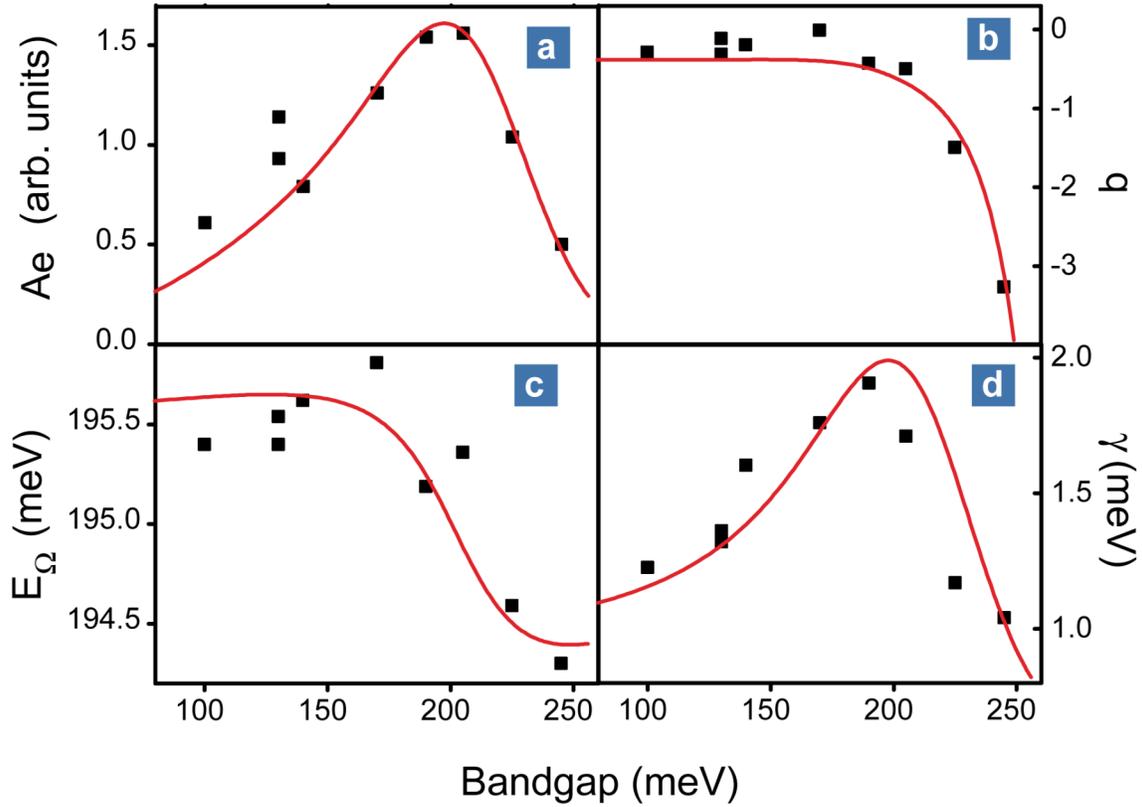

Fig. 3: Fano parameters as a function of tunable bandgap in undoped bilayer graphene. Symbols are experimental values and red lines are theoretical predictions. (a) $A_e$ describes the electronic infrared absorption at phonon frequency $E_\Omega$. It reaches maximum as the bandgap $\Delta$ comes close to $E_\Omega$. (b) Fano parameter q describes the relative dipole strengths of phonon and electronic infrared transitions. The optical phonon acquires its dipole through field-induced symmetry breaking and increases monotonically, while the electronic dipole moment reaches its maximum at $\Delta=E_\Omega$. Consequently, |q| increases dramatically at $\Delta>E_\Omega$. q is negative because coupling to electronic transitions with energy larger than $\Delta$ is stronger. (c) Phonon vibration frequency decreases at larger gap energies. This phonon softening also arises from stronger coupling to electronic transitions with energy larger than $\Delta$. (d) Phonon linewidth is broadened by the electron-phonon coupling. The linewidth reaches a maximum at $\Delta=E_\Omega$ when the electron-phonon interactions are the strongest.

Supplementary Material:

**1. Determine the gate induced bandgap and carrier concentration.**

The field induced bandgap and carrier concentration in bilayer graphene are independently varied by controlling the both the top and bottom displacement electrical fields $D_t$ and $D_b$ with the corresponding gates. The relation between the displacement electrical field and gate voltages is described by $D_{b(t)} = +(-)\varepsilon_{b(t)}(V_{b(t)} - V^0_{b(t)})/d_{b(t)}$, where the top gate dielectric layer has a dielectric constant $\varepsilon_t = 7.5$ for amorphous $Al_2O_3$ and a thickness $d_t = 80$ nm, and where the bottom gate dielectric layer has a dielectric constant $\varepsilon_b = 3.9$ for $SiO_2$ and a thickness $d_b = 285$ nm. $V^0_{b(t)}$ is the effective bottom (top) offset voltage due to initial environment induced carrier doping, which is determined through electrical transport characterization following Ref. [21] and [16].

The carrier concentration in bilayer graphene is set by the discontinuity of the field $\delta D = D_b - D_t$. Following Maxwell's equation, we have carrier concentration of $n = \varepsilon_0 \cdot \delta D$, which corresponds to $5.6 \times 10^{12} cm^{-2}$ for $\delta D$ of 1 V/nm.

The tunable bandgap $\Delta$ is determined by the average displacement electrical field $\bar{D} = (D_b + D_t)/2$. Self-consistent tight binding calculation gives an accurate description of the bandgap energy as a function of $\bar{D}$.[16,28] For $\bar{D}$ <3 V/nm, it can be well approximated by the expression $\Delta = 84 \times \bar{D} + 34.7 \times \sin(\bar{D})$, with $\Delta$ expressed in *meV* and $\bar{D}$ in *V/nm*.

**2. Detailed calculations of the many body Fano system in bilayer graphene.**

We will focus on the tunable Fano resonance in bilayer graphene at different induced bandgap but with zero carrier doping. The Hamiltonian describing low energy excitations in this many body system with one dominant active phonon mode is

$$H = E_\Omega^0 a^+ a + \sum_{\alpha,k} E_k d^+_{\alpha\bar{k}} d_{\alpha\bar{k}} + N^{-1/2} \sum_{\alpha k} M_{\alpha\bar{k}} (a^+ d_{\alpha\bar{k}} + a\, d^+_{\alpha\bar{k}}),$$

where the elementary excitations are described by the phonon creation operator $a^+$ and electron-hole pair creation operator $d^+_{\bar{k}} = c^+_{c,\bar{k}} c_{v,\bar{k}}$, $E_\Omega^0$ is the bare phonon energy, $\alpha$ denotes the electron spin and valley degrees of freedom, $E_k = E_{c,\bar{k}} - E_{v,\bar{k}}$ is the conduction and valence electron energy difference at wavevector $\bar{k}$, $N$ is the number of electronic states, and $M_{\alpha\bar{k}}$ is the electron-phonon coupling matrix element. We considered here only transitions with zero momentum change, since these are the states that couple to infrared photons which have negligible momentum. The whole system is taken to be in a finite box of volume $V$ so that the values for $\bar{k}$ and $E$ are discrete; the matrix element $M_{\alpha\bar{k}}$ is evaluated with the electron and phonon wavefunctions normalized to one in a single bilayer graphene unit cell.

The energy dispersion $E_{c,\bar{k}}$ and $E_{v,\bar{k}}$ [5,7,29] and the electron-phonon coupling matrix $M_{\alpha\bar{k}}$ [6] are calculated using self-consistent tight-binding model. In bilayer graphene two G-mode optical phonons, the symmetric and antisymmetric modes, are present. Only the symmetric mode (Fig. 1a in the main text) couples to electron-hole excitations when n ≈ 0 (due to electron-hole symmetry) and needs to be considered in our Fano system [6].

Since there is only one phonon which is relevant in the problem, to deal with the large number of degenerate electron-hole pair states, we introduce a particular state composed of superposition of electron-hole excitations of energy $E$ as

$$d^+_E = N_E^{-1/2} V_E^{-1} \sum_{\alpha k} \delta_{E_k,E} M_{\alpha\bar{k}} d^+_{\alpha\bar{k}}, \quad \text{(S1)}$$

where $V_E = \sqrt{N_E^{-1} \sum_{\alpha k} \delta_{E_{\vec{k}},E} |M_{\alpha \vec{k}}|^2}$ is an average coupling strength and $N_E = \sum_{\alpha k} \delta_{E_{\vec{k}},E}$ is the number of electronic states with energy $E$. Only this superposition of electron-hole pair states couples to the phonon vibration, and any other electron-hole pair state orthogonal to it has zero phonon coupling. With this construction, the many-body Hamiltonian in bilayer graphene can be rewritten in the form

$$H = \left( E_\Omega^0 a^+ a + \sum_E E d_E^+ d_E + \tilde{N}^{-1/2} \sum_E V_E \left( a^+ d_E + d_E^+ a \right) \right) + \left( \sum_{\alpha \vec{k}} E_{\alpha \vec{k}} d_{\alpha \vec{k}}^+ d_{\alpha \vec{k}} - \sum_E E d_E^+ d_E \right). \quad (S2)$$

Here, $\tilde{N}$ is the number of different $d_E^+|0\rangle$ states defined in Eq. (S1), i.e., having different energies. The first part of the Hamiltonian describes the phonon and the specific electron-hole excitations $d_E^+|0\rangle$ that couple to the phonon. It is responsible for all the Fano behavior in bilayer graphene. The second part describes all the other electron-hole pair excitations which are decoupled to the phonon. These electron-hole pair states can still be IR active and contribute to the optical absorption. But they produce only a relatively broad and continuous absorption background which does not interfere with the discrete phonon resonance.

To gain physical insights of the coupled phonon and electron-hole pairs system, i.e., the terms in the first parenthesis in Eq. (S2), we use second-order perturbation theory. [We have confirmed that the results from second-order perturbation theory are in quantitative agreement with those obtained from a full diagonalization of the Hamiltonian shown in Eq. (S2).] Under a canonical transformation, the Fano resonance part of the many-body Hamiltonian can be diagonalized to $H = \sum_E E \cdot h_E^+ h_E$ to second order in $|V_E|$ with the introduction of the hybrid phonon-exciton excitation creation operator

$$h_E^+ = \sin\theta_E \cdot (\frac{V_E}{\gamma} a^+ + \sum_{E'} \frac{V_E V_{E'}}{\gamma} \operatorname{Re} G^0_{EE'} \cdot d^+_{E'}) + \cos\theta_E \cdot d^+_E.$$ Here $\tan\theta_E \equiv \frac{\gamma}{E - E_\Omega}$ and

$G^0_{EE'} = \frac{1}{E - E' - i\gamma_e}$ is the bare electron-pair state Green's function [1]. The first term in $h^+$ corresponds to the renormalized discrete transition. It describes the phonon 'dressed' by its couplings to the off-resonance electron-hole pair excitations. This virtual electron-hole cloud gives the phonon its infrared activity. The renormalized phonon center frequency $E_\Omega$ and broadening $\gamma$ is given by $E_\Omega = E^0_\Omega + \sum_{E'} |V_{E'}|^2 \operatorname{Re} G_{E^0_\Omega E'}$ and

$$\gamma = \gamma_0 + \sum_{E'} |V_{E'}|^2 \operatorname{Im} G_{E^0_\Omega E'}.$$

With the hybrid phonon-exciton states being eigenstates of bilayer graphene excitations, the infrared absorption spectrum can be obtained by evaluating the optical transition matrix element of this hybrid phonon-exciton excitations. The interference between the phonon (first term of $h^+$) and the exciton (second term of $h^+$) components produces the observed Fano lineshapes. This calculation quantitatively reproduces the observed gate-tunable Fano resonances in bilayer graphene, and the comparison of the experimental and theoretical results are shown in Fig. 3 in the main text.

Another important ingredient in describing the Fano resonance is the parameter $q$ (see the main text) calculated, in our case, by 

$$q = \frac{\sum_E \operatorname{Re} \frac{1}{E^0_\Omega - E - i\gamma_e} \sum_{\alpha k} \delta_{E_{\bar{k}},E} M^*_{\alpha\bar{k}} \langle 0|d_{\alpha\bar{k}} \hat{T}|0\rangle}{\sum_E \operatorname{Im} \frac{1}{E^0_\Omega - E - i\gamma_e} \sum_{\alpha k} \delta_{E_{\bar{k}},E} M^*_{\alpha\bar{k}} \langle 0|d_{\alpha\bar{k}} \hat{T}|0\rangle}$$

where $\hat{T}$ is the operator for electron-hole pair generation through photon absorption. To account for the finite lifetime and inhomogeneous broadening of the electron-hole pair excitations, we have set the electron-hole transition width $\gamma_e$ to be 40 meV.

21